\documentclass[12pt]{article}
\usepackage{float} 
\usepackage{hyperref}

\usepackage{caption}
\usepackage{newfloat} 

\usepackage{newtxtext,newtxmath}

\usepackage{graphicx}

\usepackage[letterpaper,margin=1in]{geometry}

\renewenvironment{abstract}
	{\quotation}
	{\endquotation}

\date{}

\makeatletter
\renewcommand{\fnum@figure}{\textbf{Figure \thefigure}}
\renewcommand{\fnum@table}{\textbf{Table \thetable}}
\makeatother

\usepackage{scicite}

\usepackage{url}

\def\scititle{Modeling Extensive Defects in Metals through Classical Potential-Guided Sampling and Automated Configuration Reconstruction}
\title{\bfseries \boldmath \scititle}

\author{%
    Fei Shuang$^{1\ast}$,
    Kai Liu$^{1}$,
    Yucheng Ji$^{1,2}$, \\
    Wei Gao$^{3,4}$,
    Luca Laurenti$^{5}$,
    Poulumi Dey$^{1\ast}$\and
\begin{tabular}{@{}p{\textwidth}@{}}
    \raggedright 
    \small $^{1}$Department of Materials Science and Engineering, Faculty of Mechanical Engineering, Delft University of Technology, Mekelweg 2, Delft, 2628 CD, The Netherlands.\\
    \small $^{2}$Beijing Advanced Innovation Center for Materials Genome Engineering, National Materials Corrosion and Protection Data Center, Institute for Advanced Materials and Technology, University of Science and Technology Beijing, Beijing 100083, China.\\
    \small $^{3}$J. Mike Walker’66 Department of Mechanical Engineering, Texas A\&M University, College Station, TX 77843, United States.\\
    \small $^{4}$Department of Materials Science \& Engineering, Texas A\&M University, College Station, TX 77843, United States.\\
    \small $^{5}$Delft Centre of System and Control (DCSC), Faculty of Mechanical Engineering, Delft University of Technology, Mekelweg 2, Delft, 2628 CD, The Netherlands.\\
    \small$^\ast$Corresponding authors. Emails: P.dey@tudelft.nl; F.Shuang@tudelft.nl
\end{tabular}
}

\begin{document} 

\maketitle


\newpage


\section*{Abstract}
\begin{abstract} \bfseries \boldmath

Extended defects such as dislocation networks and general grain boundaries are ubiquitous in metals, and accurately modeling these extensive defects is crucial for understanding their deformation mechanisms. Existing machine learning interatomic potentials (MLIPs) often fall short in adequately describing these defects, as their significant characteristic sizes exceed the computational limits of first-principles calculations. In this study, we address these challenges by establishing a comprehensive defect genome through empirical interatomic potential-guided sampling. To further enable accurate first-principles calculations on this defect genome, we have developed an automated configuration reconstruction technique. This method transforms defect atomic clusters into periodic configurations through precise atom insertion, utilizing Grand Canonical Monte Carlo simulations. These strategies enable the development of highly accurate and transferable MLIPs for modeling extensive defects in metals. Using body-centered cubic tungsten as a model system, we develop an MLIP that reveals unique plastic mechanisms in simulations of nanoindentation. This framework not only improves the modeling accuracy of extensive defects in crystalline materials but also establishes a robust foundation for further advancement of MLIP development through the strategic use of defect genomes.

\end{abstract}

\newpage

\noindent
\section*{INTRODUCTION}
In the intersecting realms of computational chemistry, materials science, and mechanics, machine learning has made substantial strides, particularly through the development of machine learning interatomic potentials (MLIPs). These innovative tools have transformed atomic-scale simulations by facilitating the accurate modeling of complex materials behavior with quantum-level accuracy \cite{Mishin2021,Deringer2019,Unke2021,Wang2024,Fedik2022,Deringer2021}. A diverse array of MLIPs leveraging unique descriptors has emerged. For instance, Neural Network Potentials (NNPs) employ atom-centered symmetry functions (ACSF) \cite{Behler2011}, while Gaussian Approximation Potentials (GAP) utilize the smooth overlap of atomic positions (SOAP) \cite{GAP-2010}, and Spectral Neighbor Analysis Potentials (SNAP) depend on bispectrum components \cite{Thompson2015}. Additionally, both Moment Tensor Potential (MTP) \cite{Shapeev2016} and Atomic Cluster Expansion (ACE) \cite{Drautz2019} have found extensive applications across various materials, from metals and alloys to 2D materials and complex systems like carbon and silicon \cite{Yin2021,Mortazavi2021,Lysogorskiy2021}. A Pareto analysis reveals that while GAP achieves the highest accuracy, it lacks computational efficiency \cite{Zuo2020}. In contrast, MTP and ACE present a more balanced profile of accuracy and computational demand, with ACE being notably faster than MTP \cite{Lysogorskiy2021,Bochkarev2022,Dusson2022}.

An essential element in developing MLIPs is constructing a good training database that adequately covers the configuration space relevant to the intended simulations. Typically, this database is curated by domain knowledge, such as ground-state structures, structures deformed under various elastic strains, ab initio molecular dynamics (AIMD) configurations at different temperatures, and defects such as vacancy, interstitial, crack tip and dislocation core \cite{Si-GAP-2018}. Advancements in machine learning have driven the adoption of on-the-fly active learning (OTF-AL) \cite{Podryabinkin2017,Zhang2019,Vandermause2020}, a sophisticated approach that accelerates the development of MLIPs by efficiently selecting new configurations exclusively for necessary DFT calculations. The effectiveness of OTF-AL strategies hinges on the ability to calculate similarities among configurations by comparing their local atomic environments (LAEs). MTP with OTF-AL, utilizing the D-optimality criterion and the MaxVol algorithm, exemplifies one of the most widely employed MLIPs, which has been used in numerous studies \cite{Novikov2021,Podryabinkin2023}. Other active learning strategies, such as uncertainty-driven dynamics \cite{Kulichenko2023} and hyperactive learning \cite{npj-2023-AL}, also significantly contribute to efficiently sampling the configuration space.

Two primary challenges persist in the construction of databases for the development of MLIPs. The first challenge is the comprehensive coverage of defect. For metals, it is relatively straightforward to train MLIPs that can reproduce lattice constants, elastic constants, vacancy formation energy, and energies of simple grain boundaries. However, it remains unclear whether existing MLIPs can effectively model extended defects and their interactions, even if some of them are deemed for general-purpose \cite{Si-GAP-2018,Erhard2024,Deringer2020,Pun2020,Meng2021}. This uncertainty primarily stems from the lack of quantification for unknown defects. Although OTF-AL has been proposed to accelerate the development of MLIPs, it is insufficient to sample comprehensive defects in every loading condition. Qi et al. suggested that an ideal strategy would be to efficiently generate and sample configuration space before conducting any DFT calculations \cite{Qi2024}. They used M3GNet \cite{Chen2022} as the engine for effective and efficient database generation for atomic hydrogen diffusion in titanium hydrides system. This strategy however is unsuitable for modeling extended defects in metals due to the low accuracy and computational efficiency of universal MLIPs in large-scale simulations.

The second challenge involves the accurate modeling of extended defects. For instance, while current MLIPs can handle local defects such as vacancies, screw dislocations, and simple grain boundaries, they often exclude extended defects such as dislocation nucleation and multiplication, general grain boundaries (GBs), and interactions among them \cite{Freitas2022}. This exclusion is due to the inability of these extended defects to fit within the small periodic configurations typically used in DFT calculations. One workaround has been to extract a non-periodic atomic cluster with a spherical shape from large-scale simulations. To restore periodic boundary conditions required for plane-wave DFT calculations, a sufficiently large vacuum layer is added to prevent image interaction, as implemented in the MLIP-3 package \cite{Podryabinkin2023}. This strategy has been applied in nanoindentation simulations \cite{Podryabinkin2022} and elastic constant calculations of polycrystals \cite{Jalolov2024} for diamond. Nevertheless, this approach can introduce irrelevant free surfaces or other boundaries \cite{Lysogorskiy2023}. To overcome these limitations, new strategies have been developed to maintain periodic boundary conditions without additional vacuum layers. For instance, Zhang et al. created a periodic crack-tip DFT cell through duplication and rotation operations \cite{Zhang2023}. Hodapp et al. utilized a screw dislocation in BCC W as an example to construct a periodic configuration \cite{Hodapp2020}, while Mismetti et al. designed periodic configurations containing leading and trailing parts of edge dislocation in face-centered cubic aluminum (FCC Al) \cite{Mismetti2024}. However, these methods fall short in handling aforementioned extended defects.

In this study, we introduce a generalized framework for developing MLIPs specifically designed to model extensive defects in metals. This framework integrates two key components: (1) Empirical Interatomic Potentials-Guided Sampling (EIP-GS), and (2) Periodic Configuration Construction via Grand Canonical Monte Carlo simulations (PCC-GCMC). These components are crucial for overcoming the challenges previously discussed. The EIP-GS method is designed to enhance defect sampling capabilities during large-scale simulations under a wide range of loading conditions. The PCC-GCMC technique, on the other hand, converts non-periodic atomic clusters of selected defects into periodic configurations without the need for vacuum. This conversion is essential for ensuring that these configurations are compatible with standard plane-wave DFT calculations. Focusing on body-centered cubic tungsten (BCC W), our analysis underscores the critical role of the EIP-GS and PCC-GCMC techniques in enhancing the predictive accuracy and reliability of MLIPs for large-scale simulations. Ultimately, we have developed a set of MLIPs optimized for BCC W, capable of simulating a wide range of plastic deformations. We demonstrate the capabilities of our MLIPs through nanoindentation simulations on single crystals. These tests reveal new deformation mechanisms not previously documented, underscoring the robustness and adaptability of our MLIPs in handling complex simulation scenarios.

\section*{RESULTS}
\subsection*{MLIP development framework for modeling extensive defects}
In this study, we introduce an generalized framework designed to develop versatile MLIPs for accurately modeling extensive defects in metals, as depicted in Fig. \ref{fig1}. This framework is implemented through a sequential three-step process. Initially, a comprehensive pool of defect structures is generated using large-scale simulations based on well-established EIPs, notably the Embedded-Atom Method (EAM) potentials for metals. These potentials have been instrumental for over three decades, enabling the rapid simulation of complex defects across millions of atoms, albeit with lower accuracy compared to DFT. These capabilities are indispensable for simulating a broad spectrum of deformation scenarios, encompassing general GBs in polycrystals, as well as compression, tension, shear, and nanoindentation tests in both polycrystalline and single-crystal systems (Fig. \ref{fig1}a). For a particular metal (BCC W in this study), we employ EAM potentials to generate a diverse set of LAEs for defect, addressing phenomena such as grain boundary relaxation and deformation, dislocation nucleation from material inside and exposed surfaces, dislocation multiplication, and complex dislocation-grain boundary interactions. Subsequently, the D-optimality criterion-base algorithm implemented in MLIP-3 \cite{Podryabinkin2023} is used to select representative LAEs based on the local atomic neighborhoods. This strategy meticulously evaluates the extensive pool of generated configurations against an established basic dataset (obtained through domain knowledge, as detailed in Supplementary Note 1), enabling the precise selection of the most representative LAEs. This process yields distinct atomic clusters with typical size of 100-200 atoms. For these clusters, vacuum layers of 8 \AA{} thickness are added in all three dimensions of the simulation box to minimize image interaction. We develop an MLIP, named C-MLIP-0, combining the basic dataset with these cluster configurations.

In the second step, our PCC-GCMC strategy is employed to convert all clusters from the first step into periodic configurations as shown in Fig. \ref{fig1}b. For each atomic cluster, we begin by initializing the simulation box size slightly larger than the atomic coordinates in all directions. Subsequently, new atoms (depicted in blue in Fig. \ref{fig1}b) are inserted into the box via GCMC simulations, continuing until no additional atoms can be accommodated. EAM potentials are used in GCMC simulations. During this insertion phase, the core atoms (shown in red) remain fixed to preserve the original LAE of the center atom. After the insertion is complete, both the coordinates of the newly inserted blue atoms and the box size are relaxed simultaneously. This step is crucial to prevent highly distorted LAEs among the newly inserted atoms. This procedure is iteratively repeated, with the box size gradually increased to enable the insertion of atoms into lower energy states. By selecting only the lowest-energy configurations from PCC-GCMC, we ensure that all LAEs in these configurations are physically pertinent to the central LAE. The resulting new DFT dataset, combined with the basic dataset, is utilized to train a foundational model, termed P-MLIP-0, designed to be suitable for simulating common defects. 

The final stage, Step 3, is the MLIP refinement phase, as illustrated in Fig. \ref{fig1}c. For specific applications not addressed in the initial EIP-GS of Step 1, such as crack propagation and radiation damage, further EIP-GS and PCC-GCMC operations are required, utilizing appropriate EIPs. A limitation of EIP-GS is its potential bias in generating necessary LAES due to the inherent bias of EIPs. Consequently, new LAEs may be generated during MLIP-based large-scale simulations. To ensure the transferability of MLIPs across diverse applications, we implement the traditional OTF-AL process here for specific cases. All generated atomic clusters are converted into periodic configurations via PCC-GCMC and reincorporated into the training dataset. This process is repeated until no new LAEs are produced by OTF-AL. The final version of our MLIP, named P-MLIP-1, represents the culmination of our development efforts and is designed to robustly model a wide range of defect phenomena.

\subsection*{Significance of PCC-GCMC}

In this section, we discuss the significance of PCC-GCMC. To demonstrate the functionality of PCC-GCMC, we employ the example of a screw dislocation core in BCC W, as illustrated in Fig. \ref{fig2}a. The atomic cluster includes a screw dislocation core with three compact core atoms, labeled 1-3. This configuration maintains periodic boundary conditions along the $z$-direction, with free boundary conditions along the $x$ and $y$ directions. The PCC-GCMC process is repeated with varying box dimensions, $\Delta_x$ and $\Delta_y$, from 1 to 10 \AA. Fig. \ref{fig2}b shows the average potential energies of each configuration constructed via PCC-GCMC. Notably, configurations with potential energies below -8.8 eV/atom, depicted in blue, approach the cohesive energy of -8.9 eV/atom which are nearly close to a perfect BCC W. Fig. \ref{fig2}c highlights two representative configurations. In the top panel, we observe a periodic configuration featuring a new non-compact core crossing the periodic boundary along the $y$-axis dislocation, with the remainder of the atoms conforming to the perfect BCC lattice structure, excluding the original screw core. The bottom panel reveals a periodic configuration that includes a screw dislocation dipole, akin to typical screw dislocation dipole configurations with a tilted box. Both configurations depicted in Fig. \ref{fig2}c are generated entirely through automated processes. All newly inserted atoms either form a pure BCC lattice or constitute relevant LAEs, making them suitable for MLIP training. 

We generate both the cluster dataset and the periodic dataset following the procedure outlined in Fig. \ref{fig1}a. All large-scale simulations are driven using the EAM-Zhou potential \cite{Zhou2004}. We then explore the differences between the cluster dataset shown in Fig. \ref{fig1}a and the periodic dataset from Fig. \ref{fig1}b, alongside the basic dataset informed by domain knowledge. This examination is conducted by analyzing the difference in the SOAP descriptor \cite{SOAP-2013} with parameters set as $r_\text{cut}$ = 5 \AA, $n_\text{max}$ = 12, and $l_\text{max}$ = 10 \cite{Himanen2020}. The results are visualized in Fig. \ref{fig2}d, where the first two principal components (PCs) are plotted against each other. In the basic dataset, the majority of atomic environments exhibit PC1 values less than 0, with some data points extending into a long arm, representing surface atoms. The cluster dataset, conversely, divides into two groups: a smaller group that aligns closely with the main portion of the basic dataset and a larger group that positions itself above the extended arm of the basic dataset. This larger group primarily comprises surface atoms, a logical outcome given their proximity to the free surface. The smaller group represents target atoms located at the center of each cluster configuration, which correspond to the LAEs of defects. The size disparity between these groups indicates that only a small fraction of LAEs directly contribute to target defects, while the majority, being surface atoms, do not inform the force and energy calculations of non-surface atoms. 

Unlike the cluster dataset, the periodic dataset forms a single cohesive group containing all target atoms from the cluster dataset. There is an overlapping region between the periodic and basic datasets, marked by a dashed line, suggesting shared common LAEs. This overlap signifies that the periodic configurations created via our PCC-GCMC method exhibit characteristics similar to those identified in Ref \cite{Kong2023}, with data points in this region likely corresponding to the perfect BCC lattice. Additionally, other data points in the periodic dataset represent inserted atoms from the GCMC process and atoms close to the surfaces in original clusters. Although these LAEs are not explicitly defined in the initial EIP-GS (such as non-compact screw core discovered in Fig. \ref{fig2}c), they may prove instrumental in exploring other types of unknown plastic deformations, which we will discuss further in the following section. Another interesting observation is that the LAEs of defects are not included in the basic dataset (\ref{fig2}d), indicating that the existing MLIPs obtained through domain knowledge are unsuitable for modeling general defects.

Next, we proceed to train C-MLIP-0 and P-MLIP-0, utilizing the cluster dataset (C) and the periodic dataset (P), respectively, each integrated with the basic dataset (B). MTP is used for training. The training performance for these models is assessed by comparing the root mean square error (RMSE) of configuration energies, as presented in Fig. \ref{fig2}(e). It is consistently observed that the B+P combination outperforms B+C across various MTP levels. This superior performance of B+P can be attributed to the more straightforward nature of the dataset, which is devoid of the complex, irrelevant LAEs like surface atoms found in the cluster dataset (illustrated in Fig. \ref{fig2}d). This finding underscores the first significant advantage of employing the periodic dataset in MLIP training: the elimination of irrelevant LAEs enhances the training process, leading to more accurate and reliable MLIPs.

\subsection*{Uncertainty quantification across diverse plastic deformations}
Our MLIP-0 models, i.e., C-MLIP-0 and P-MLIP-0, are designed to accommodate a wide range of defect simulations including dislocations, GBs, dislocation nucleation and multiplication, and dislocation-GB interactions. To verify the transferability of our MLIP-0 models, we utilize per-atom uncertainty quantification available in MLIP-3 \cite{Podryabinkin2023} (the extrapolation grade $\gamma$) to evaluate various defects in BCC W. Values of $\gamma$ ranging from 0 to 1 signify interpolation, whereas values exceeding 1 imply extrapolation. Figs \ref{fig3}a-c display three common types of dislocation loops encountered in the plastic deformation of BCC metals: a $\langle100\rangle$ interstitial loop, a $\langle111\rangle$ glide loop, and a $\langle111\rangle$ climb loop. Given that the performance outcomes of the basic dataset combined with cluster one (B+C) and the basic dataset with periodic one (B+P) are comparable, we present only the results for B+C for clarity. Notably, using the basic dataset (B), all atoms within the $\langle100\rangle$ interstitial loop and those near the edge segments of the $\langle111\rangle$ glide loop, along with some atoms in the $\langle111\rangle$ climb loop, exhibit $\gamma$ values significantly greater than 1, indicating a higher level of uncertainty (left panels of Fig. \ref{fig3}a-c). In contrast, all atoms within these dislocation loops, when analyzed using the B+C dataset, show $\gamma$ values less than 1. This demonstrates our MLIP-0 models' enhanced effectiveness in capturing complex dislocations, with energies and forces of all LAEs being obtainable through interpolation.

Additionally, we assess the transferability of our models regarding general GBs in polycrystals and their temperature-induced evolution. We calculate the extrapolation grade $\gamma$ for all the atoms in a polycrystal relaxed by a MEAM potential \cite{Hiremath2022} at 300 K and 2000 K. It should be noted that our EIP-GS in the initial EIP-GS only use EAM-Zhou potential \cite{Zhou2004}, such that the GBs obtained by MEAM may have different LAEs. Fig. \ref{fig3}d illustrates the polycrystal at 300 K, with each atom color-coded according to $\gamma$ values obtained from B+C. Notably, all atoms in this representation exhibit $\gamma$ values less than 1, indicating low uncertainty and high confidence in the predictions. Figs \ref{fig3}e and f display histogram plots of the $\gamma$ values for this polycrystal calculated using different datasets: the basic dataset (B), a dataset from a Gaussian Approximation Potential (GAP, see the Supplementary Note 1 for the details) \cite{W-GAP-2014}, B+C, and B+P at 300 K and 2000 K. The analysis reveals that both datasets B and GAP predict $\gamma$ values greater than 2 for atoms at both 300 K and 2000 K, indicating higher uncertainty. This is particularly pronounced for the dataset B. In contrast, $\gamma$ values calculated by datasets B+C and B+P are consistently below 1 at 300 K. However, at 2000 K, there are a few exceptions with values of $\gamma$ slightly higher than 2. This indicates that although the models perform well at lower temperatures, at higher temperatures the uncertainty escalates but stays manageable. These results underscore the high transferability and effectiveness of generated defect genome in simulating the behavior of general GB and the thermodynamics of polycrystals.

We further evaluate the uncertainty quantification of our defect datasets during a continuous nanoindentation simulation, illustrated in Figs \ref{fig3}(g-i). Figs \ref{fig3}(g, h) display the distribution of defect atoms color-coded by $\gamma$ values at the maximum depth (4 nm), as calculated using B and B+C, respectively. For the case of B, several regions exhibit $\gamma$ values above 1, particularly in regions involving dislocation interactions and surface regions, as indicated by red arrows in Fig. \ref{fig3}(g). In contrast, all defect atoms analyzed with B+C demonstrate $\gamma$ values lower than 1 (Fig. \ref{fig3}(h)), indicating significantly lower uncertainty. Additionally, we calculate the maximum $\gamma$ values ($\gamma_\text{max}$) among all defect atoms for each snapshot throughout the nanoindentation process. Comparisons of $\gamma_\text{max}$ for B, GAP, B+C, and B+P reveal that B and GAP consistently show $\gamma_\text{max}$ values above 2, reflecting high uncertainty throughout the nanoindentation process. Conversely, both B+C and B+P maintain $\gamma$ values around 1, with B+C exhibiting slightly lower $\gamma$ values than B+P. This difference is attributed to B+C having a higher proportion of surface atoms, which are prevalent in open-surface nanoindentation simulations. These results indicate that our defect pool encompasses all the plastic LAEs during nanoindentation, including dislocation nucleation from open surfaces and complex multiplication beneath the indenter. It should be noted that datasets B+C and B+P exhibit comparable performance in uncertainty quantification analysis in Fig. \ref{fig3}. This is because the key LAEs for these defects are included in the core atoms of both datasets. However, this does not mean that dataset P, constructed using PCC-GCMC, is unnecessary. In the following section, we will show that dataset P outperforms dataset B in predicting unseen LAEs.

Furthermore, we compare the LAEs in the existing GAP dataset with those in B+P and B+C, as detailed in Supplementary Note 1. Our analysis shows that the GAP dataset and its corresponding MLIP are inadequate for effectively modeling general defects in BCC W. Taken together with the results shown in Fig. \ref{fig3}, these insights emphasize the critical role of EIP-GS as an advanced sampling method, crucial for ensuring data diversity and comprehensive coverage.. 

\subsection*{Direct Validation Through Comparison with DFT Calculations}

 To further elucidate the superiority of PCC-GCMC and the resultant P-MLIP-0, we undertake additional validation tasks by comparing the energy and forces predicted by our MLIP-0 models against new DFT calculations for various GBs relaxed at temperatures of 300 K, 900 K, and 2000 K, as well as under severe compression conditions. The stable structures of these GBs are referenced from an established study in Ref \cite{GB-2020}. The GB structures and comparative results for energy and force between MLIPs and DFT are displayed in Fig. \ref{fig4}. Nine distinct GBs are considered, which transform into FCC lattices, hexagonal close-packed (HCP) lattices, and even an amorphous phase under severe compression, as shown in Fig. \ref{fig4}a. Observations reveal that both C-MLIP-0 (trained by B+C) and P-MLIP-0 (trained by B+P) consistently outperform B-MLIP-0 (trained solely by B) across all validation scenarios. An exception is noted in the energy comparison at 2000 K, where B-MLIP-0 records a lower error than C-MLIP-0 and P-MLIP-0 (6.3 vs 12.2 and 7.2 meV/atom). However, the force RMSE for B-MLIP-0 is significantly higher compared to C-MLIP-0 and P-MLIP-0 (199.3 vs 172.2 and 154.4 meV/\AA). These results suggest that incorporating either cluster (C) or periodic (P) datasets enhances the accuracy. Notably, P-MLIP-0 consistently surpasses C-MLIP-0 in both energy and force accuracy, particularly under compression conditions (6.7 vs 10.1 meV/atom for energy; 150.8 vs 173.7 meV/\AA{} for force). This superior performance can be attributed to the inclusion of additional atoms by PCC-GCMC in Fig. \ref{fig2}d, which introduces relevant LAEs absent in B+C, and thus better aligns with the new validation tasks in Fig. \ref{fig4}b. Combining the training performance illustrated in Fig. \ref{fig2}e, we conclude that B+P offers superior training outcomes and higher transferability than B+C, thereby validating its enhanced efficacy in modeling complex defects. 

\subsection*{MLIP Refinement}
It should be noted that all LAEs discussed above are generated using the EAM-Zhou potential \cite{Zhou2004}. The applicability of the MLIP-0 models developed based on these LAEs for modeling extended defects in large-scale simulations, particularly within MLIP simulations, remains uncertain. To address this issue, we employ the traditional on-the-fly AL (OTF-AL) implemented in MLIP-3 \cite{Podryabinkin2023}. This process allows us to monitor the extrapolation grade $\gamma$ during MLIP-based simulations and to identify unknown LAEs. These activities correspond to Step 3 in Fig. \ref{fig1}c. 

We focus on the high-temperature relaxation of a polycrystal. Fig. \ref{fig3}e shows that our dataset, constructed via the initial EIP-GS, includes all LAEs in a polycrystal relaxed by the MEAM potential at 300 K. However, Fig. \ref{fig3}f reveals that these LAEs do not fully represent those at 2000 K, making EIP-GS ineffective for high temperatures. To address this, we use OTF-AL to model polycrystal relaxation at 300 K, 900 K, 1200 K, and 2000 K, starting with the dataset B+P. In the first AL loop, no new configurations are generated at 300 K and 900 K. However, the simulation at 1200 K stops due to atom loss, leading to the selection of 21 new configurations with $\gamma>2$. These are converted into periodic configurations via PCC-GCMC and added to the training database to develop an updated MLIP. In the next loop, relaxation simulations proceed smoothly across all temperatures, with 25 additional configurations identified at 2000 K. This ensures a comprehensive defect genome for modeling general GBs in random polycrystals, all embedded in periodic configurations via PCC-GCMC. These configurations form the dataset OTF-AL-GB. To assess the uncertainty of our MLIP trained on B+P+OTF-AL-GB, we use ensemble learning (detailed in the Supplementary Note 2) to evaluate a polycrystal relaxed at 300 K and 2000 K (Fig. \ref{fig5}). The results show significantly reduced uncertainty, with most force standard deviations ($\sigma_z$) below 0.2 eV/\AA{}, comparable to MLIP force prediction errors. This confirms the high accuracy of the new MLIP in modeling general GBs.

Additionally, we refine our MLIP model by incorporating crack propagation, as detailed in Supplementary Note 3. Our approach effectively captures the local atomic environments (LAEs) associated with various fracture mechanisms. Finally, we consolidate all datasets in Table \ref{TableS1} and train a series of MTPs at multiple levels, alongside a complex ACE potential. Due to ACE's exceptional training accuracy, predictive capability for fundamental properties of BCC W, and computational efficiency (see the details in the Supplementary Note 4), we designate ACE as the final MLIP model for subsequent large-scale MD simulations.

\subsection*{MLIP applications in nanoindentation}
We initiate the application of our ACE potential by simulating nanoindentation in single crystal W. Although nanoindentation has been extensively studied using EAM potentials \cite{Li2002,Lee2005,Tan2023}, there are relatively few simulations employing MLIPs. This scarcity likely arises from the challenges of modeling dislocation nucleation from open surfaces and capturing complex dislocation interactions, which are difficult to simulate directly with DFT. In a recent study, a tabulated GAP potential was used for nanoindentation of single crystal W \cite{W-indent}, but the uncertainty associated with these simulations was not detailed. Meanwhile, an advanced neural network potential was specifically developed for nanoindentation in single crystal Mo \cite{Naghdi2024}. We now present the simulation results obtained with our final ACE potential and examine the uncertainty.

The simulation setup is aligned with Ref \cite{W-indent}, depicted in Fig. \ref{fig6}a. The simulation details are provided in Methods. Fig. \ref{fig6}b displays the force-depth curve, reaching a maximum depth of 3.8 nm. The corresponding microstructures at this depth are depicted in Fig. \ref{fig6}c, where all defect atoms are color-coded based on the extrapolation grade ($\gamma$). Notably, all defects exhibit $\gamma$ values lower than 1.2, affirming the low uncertainty in modeling the complex dislocation network generated by the ACE potential. The dislocation pattern reveals prismatic loops advancing along $\{111\}$ slip directions. This pattern, which includes the process of shear loop formation and cross-slip repeating beneath the indenter tip, aligns with results from previous studies using tabGAP and experimental observations \cite{W-indent}. Additionally, the slip traces and pileup, detailed in Fig. \ref{fig6}d, resonate with findings from the customized neural network potential employed for Mo nanoindentation \cite{Naghdi2024}. These results not only validate the effectiveness of the ACE in capturing intricate deformation mechanisms but also demonstrate its consistency with established models and experimental data.

Our simulations reveal a new mechanism for the onset of plasticity during nanoindentation, as analyzed using BCC Defect Analysis (BDA) \cite{BDA}. The outcomes of this analysis are illustrated in Fig. \ref{fig6}e, where nano-screw dislocations, deformation twinning, $\{110\}$ planar faults, surface features, and vacancies are identified. Specifically, the sequence of events leading to the initiation and evolution of plasticity is marked by five critical points A-E, labeled in Fig. \ref{fig6}b and detailed in Fig. \ref{fig6}e. At point A, plasticity begins with the formation of a local HCP region. The first yield, at point B, is associated with the formation of three intersecting twin boundaries (TBs). At point C, the vertical TB extends deeper into the material while the left one disappears. At point D, the vertical TB further advances, and the right TB transitions into a regular GB. Finally, at point E, the extension of the vertical TB halts. These results indicate that while deformation twinning is critical at the initial stages of nanoindentation (from B to D), dislocation nucleation and multiplication subsequently take precedence, governing the later stages of deformation (from D to E).

\section*{DISCUSSION}
Over the past decade, MLIPs have emerged as a cornerstone in computational science, significantly impacting fields such as chemistry, materials science, and mechanics \cite{Mishin2021,Deringer2019,Unke2021,Wang2024,Fedik2022,Deringer2021}. Various frameworks employing a range of atomic descriptors, including ASCF \cite{Behler2011}, SNAP \cite{Thompson2015}, moment tensor \cite{Shapeev2016}, and ACE \cite{Drautz2019}, have been developed. These frameworks have implemented diverse training strategies, from linear regression to neural networks, primarily focusing on enhancing the accuracy and computational efficiency of MLIPs. Such advancements are crucial, as they significantly improve the ability of MLIPs to model complex atomic interactions both accurately and efficiently, thereby paving the way for groundbreaking discoveries in multiple scientific domains. However, the construction of a robust training dataset is equally essential. The quality and comprehensiveness of the data used for training directly impact the effectiveness and reliability of MLIPs. Despite these developments, the scrutiny applied to most MLIPs remains insufficient. These potentials are often deployed in large-scale applications without undergoing rigorous validation across diverse conditions or against a wide spectrum of atomic configurations. Consequently, due to inadequately comprehensive training datasets, many existing MLIPs fail to accurately model general defects, thereby limiting their applicability in complex simulations.

While the MLIP-3 package facilitates OTF-AL using MTPs to tackle similar challenges \cite{Podryabinkin2023}, our integration of EIP-GS and PCC-GCMC offers substantial improvements. EIP-GS significantly reduces the demanding requirements of OTF-AL, expediting the development of MLIPs. Typically, OTF-AL involves iterative MD simulations based on MLIPs and frequent retraining \cite{Qi2024}, a process that is time-consuming, especially for rare events like diffusion and phase transitions. Moreover, retraining MLIPs, whether using linear regression or neural networks, is computationally intensive. The data in Table \ref{TableS1} highlights this efficiency. Of the 348 configurations generated, only 72 are from OTF-AL. On the other hand, the new LAEs acquired from OTF-AL are usually specific to the simulated loading conditions and may not generalize well across different scenarios. It is impractical to perform OTF-AL for every conceivable loading condition. For example, certain LAEs that occur in low strain-rate MD simulations might not emerge under high strain rates, such as the rate-dependent ductile-to-brittle failure transition observed in metal nanowires \cite{Tao2018,Ramachandramoorthy2016}. Consequently, OTF-AL from MD simulations at high strain rates cannot be used to develop MLIPs suitable for lower strain rates, as critical LAEs at lower rates remain unexplored. This strain rate limitation has been identified as a significant challenge in advancing MLIP development \cite{Mortazavi2023}. EIP-GS effectively addresses this challenge by leveraging cost-effective EIPs to conduct extensive MD simulations. Even within the same material, various EIPs exhibit distinct properties, which can produce a diverse range of LAEs under identical loading conditions, as demonstrated by the crack propagation scenarios shown in Fig. \ref{S2}. For W alone, there are 32 different EIPs available from OpenKIM \cite{Tadmor2011}. The varied characteristics of these EIPs allow us to discover a more extensive array of LAEs than those typically identified through OTF-AL.

Our PCC-GCMC method significantly refines our approach to developing MLIPs for modeling extensive defects, serving as a critical supplement to both EIP-GS and OTF-AL. This method departs from the conventional approach used in the MLIP-3 package \cite{Podryabinkin2023}, which involves adding a vacuum layer to minimize image interactions—a process that proves inefficient due to the limited use of LAEs, with most being near the surface (Fig. \ref{fig2}a). In contrast, PCC-GCMC streamlines the creation of periodic configurations that are immediately integrated into the training dataset following DFT calculations, enhancing the dataset's relevance, as discussed in Fig. \ref{fig2}. Specifically, this method avoids the use of a vacuum layer, circumventing issues with surface atoms. The introduction of atoms in this process could yield perfect lattice or related defect structures (Fig. \ref{fig2}a), or more likely, identify previously unrecognized defects that improve the transferability of the developed MLIP (Fig. \ref{fig4}).

Our approach demonstrates significant advancements over previous studies \cite{Hodapp2020,Zhang2023,Mismetti2024}. While these earlier methods are restricted to simple defects and often rely on manual operations, such as rotation and merging in Ref. \cite{Zhang2023}, they also require elastic solutions for simple defects to construct atomistic positions, as seen in Refs. \cite{Hodapp2020,Mismetti2024}. In contrast, our methodology is uniquely capable of addressing complex defects, including mixed dislocations and general GBs, where prior approaches prove inadequate. An alternative strategy for developing general-purpose MLIP datasets (RANDSPG) involves the systematic exploration of all possible space groups in random crystal structures \cite{Poul2023,Ito2024}. Although MLIPs trained on the RANDSPG dataset achieve accurate modeling of simple GBs in Mg and general GBs in BCC Fe \cite{Ito2024}, recent findings reveal substantial limitations. Specifically, high uncertainty ($\gamma > 2$) and significant force errors are observed in polycrystal deformations of BCC Fe \cite{Ito2024-2}, as reported by the same authors in Ref. \cite{Ito2024}. In comparison, our MLIPs exhibit markedly lower uncertainty ($\gamma < 1.2$) in nanoindentation (Fig. \ref{fig6}c). These results underscore the insufficiency of the RANDSPG dataset for modeling extensive defects and highlight the superior robustness of our approach.

Finally, we propose that our framework can be generalized across a diverse array of common crystalline materials to model extensive defects, as illustrated in Fig. \ref{fig7}. This framework is versatile, applicable not only to familiar structures such as FCC, BCC, HCP, and diamond cubic but also to more complex lattices like perovskite. A rich assortment of EIPs is available from resources like the NIST Interatomic Potentials Repository (NIST-IPR) \cite{NIST-IPR} and OpenKIM \cite{Tadmor2011}. These resources provide access to a variety of potentials, including EAM, MEAM, bond order potentials (BOP), Stillinger-Weber (SW) potentials, Tersoff potentials, and reactive force fields (ReaxFF), among others. OpenKIM, for instance, hosts over 644 potentials, with a significant majority being EIPs. Computational materials science has benefited greatly over the past three decades by leveraging these EIPs \cite{Becker2013}, and despite the current emphasis on MLIPs, the value of EIPs remains substantial. Although these potentials exhibit higher force and energy errors compared to DFT calculations, their consistent internal logic as physical models provides significant value. Utilizing the abundant EIPs available, our study's approach can be used to develops advanced MLIPs through three strategic initiatives: adapting EIPs from different materials but sharing the same lattice structure by merely rescaling lattice constants to generate relevant LAEs; parameterizing EIPs to customize the mechanical properties of materials \cite{Daw2023,Rajan2016}; and employing the average-atom interatomic potential for random alloys to produce an unlimited series of EIPs from basic elemental forms \cite{Varvenne2016}. Utilizing these resources, our approach enables the creation of a standard defect genome for each crystalline structure. These genomes form the cornerstone for developing MLIPs capable of accurately modeling any type of defect. This not only facilitates further research within this study but also provides a valuable tool for the broader scientific community, enhancing the predictive capabilities of computational material science.

In summary, this work introduces an advanced framework for developing MLIPs by integrating EIP-GS, PCC-GCMC, and OTF-AL. This framework enables automated, ab initio-accurate, large-scale atomistic modeling of general defects in metals. The developed MLIPs are highly transferable and are based on comprehensive datasets that facilitate simulations of complex behaviors, including general GBs in polycrystals, dislocation nucleation and interactions, dislocation-GB interactions, and crack propagation in BCC W. Rigorous uncertainty quantification analyses, employing D-optimality criterion and ensemble learning techniques, confirm the accuracy of our models in these scenarios. We have successfully applied our trained ACE potential to uncover dominant deformation mechanisms in nanoindentation, identifying deformation twinning as the primary mechanism at the onset of plasticity, and dislocation evolution as the predominant mechanism at later stages. This approach proves versatile enough to be generalized for modeling extensive defects in other BCC metals, FCC metals, and high-entropy alloys, demonstrating its broad applicability across various material systems.

\section*{MATERIALS AND METHODS}
\subsection*{First principles calculations}
We utilize the Vienna Ab initio Simulation Package (VASP) to perform first-principles calculations of all new configurations necessary for MLIP development \cite{Kresse1996}. A gradient-corrected functional in the Perdew-Burke-Ernzerhof (PBE) form is used to describe the exchange and correlation interactions \cite{Perdew1996}. Electron-ion interactions are treated within the projector-augmented-wave (PAW) method, using the standard PAW pseudopotentials provided by VASP. The energy convergence criterion is set to $10^{-6}$ eV for electronic self-consistency calculations. The plane-wave cutoff energy is chosen to be 520 eV. The KPOINTS are generated by VASPKIT \cite{VASP-KIT}, based on the Monkhorst-Pack scheme, with a consistent density of 2$\pi$ ×0.03 \AA{} across  the entire dataset. 

\subsection*{Empirical interatomic potentials}
 In our simulations, we utilize both the Embedded-Atom Model (EAM) and the Modified Embedded-Atom Model (MEAM) to perform large-scale molecular dynamics simulations of body-centered cubic tungsten (BCC W) under various loading conditions. Specifically, we employ the EAM-Zhou potential \cite{Zhou2004} for simulating the relaxation, compression, and tension of W polycrystals, as well as single-crystal compression, tension, and nanoindentation processes. Additionally, the MEAM potential referenced in Ref \cite{Hiremath2022} is applied to model crack propagation in single-crystal W. 

 \subsection*{Atomistic simulations and analysis}
We employ the Large-scale Atomic/Molecular Massively Parallel Simulator (LAMMPS) package for all the atomistic simulations \cite{Thompson2022}. For visualizing atomic configurations and post-processing results such as extracting dislocation structures, we use OVITO \cite{Stukowski2010}. Additionally, Atomsk is used to generate polycrystalline structures \cite{Hirel2015}. These packages provide a comprehensive suite of tools to support our study of the mechanical properties and plastic deformation mechanisms of BCC W.

The indentation is conducted along the [100] direction, which is free from boundary constraints, while the other two axes maintain periodic boundary conditions. The total system spans dimensions of 38 $\times$ 38 $\times$ 32 $\text{nm}^{3}$, encompassing 2,832,200 atoms. To stabilize the substrate, the bottom layers, with a thickness of 6.4 \AA, are fixed. Additionally, several layers, totaling a thickness of 2.56 nm, function as thermostat layers to dissipate heat generated during the nanoindentation process. The remaining atoms are designated as Newtonian layers, where the indentation process occurs. The indentation is performed at a velocity of 20 m/s, using a spherical indenter with a radius of 6 nm.

\subsection*{Machine learning potential development}
We employ the MLIP-2 package \cite{Novikov2021} and pacemarker \cite{Bochkarev2022} to develop MTPs and ACE potential for BCC W. MLIP-2 utilizes moment tensor descriptors and apply linear regression to train the machine learning model, enabling it to predict the energy, force, and stress of atomistic systems. Within our study, the weights of energy, force and stress are set as 1, 0.01 and 0, respectively. For ACE, we utilize the highly nonlinear potential energy function with the complex form of $E_i = \varphi + \sqrt{\varphi} + \sum_i \varphi^{f_i}$, where $E_i = \varphi + \sqrt{\varphi} + \sum_{i} \varphi^{f_i}, \quad f_i \in \left\{ \frac{1}{8}, \frac{1}{4}, \frac{3}{8}, \frac{3}{4}, \frac{7}{8}, 2 \right\}$, obtained from Ref \cite{Erhard2024}. For the expansion of the atomic properties, we used 72 basis functions with 801 parameters. As radial basis we employ Bessel functions. A $\kappa$ value of 0.01, which gives the ratio between force and energy weights value, was used during fitting. For optimisation we use the BFGS algorithm for 2000 steps. The cutoff distance is set as 5 \AA{} for both MTP and ACE.
More importantly, we utilize MLIP-3 \cite{Podryabinkin2023} to perform active learning based on the MaxVol algorithm from extensive simulations conducted with EAM or MEAM potentials. This method was originally developed for the OFT-AL of atomic environments, incorporating a cycle of MTP-based simulations, DFT calculations, and subsequent MTP retraining. In this study, to expedite the process, we exclusively utilize the EIP-GS to identify representative defects from a comprehensive pool of large-scale simulations. This approach and its implications are further elaborated in Section 3.


\section*{Data availability}
The data that support the findings of this study are available at the GitHub repository: 
\url{https://github.com/ufsf/ML-Defect-Modeling}.

\section*{Code availability}
All simulations are executed using open-source software LAMMPS. The machine learning force field was trained and validated by the MLIP package\cite{Novikov2021}. All source codes are available at the GitHub repository: \url{https://github.com/ufsf/ML-Defect-Modeling}.

\section*{Acknowledgments}
This work was sponsored by Nederlandse Organisatie voor WetenschappelijkOnderzoek (The Netherlands Organization for Scientific Research, NWO) domain Science for the use of supercomputer facilities. The authors also acknowledge the use of DelftBlue supercomputer, provided by Delft High Performance Computing Center (https://www.tudelft.nl/dhpc).

\section*{Author Contributions}
F.S. performed DFT calculations, developed the machine learning force field, performed the atomistic simulations, and wrote the first draft of the manuscript, K.L. and F.S. proposed the PCC-GCMC approach. F.S., L.L. and P.D. conceptualized the project and designed the research. F.S., K.L., Y.J., G.W., L.L. and P.D. analyzed the simulation data. All the authors contributed to the interpretation of the data.

\section*{Conflict of Interest}
The authors declare no conflict of interest.

\clearpage 

%
\bibliography{ref} 
\bibliographystyle{sciencemag}

%
%
%
%
%
%


\newpage
\section*{Figures}

\begin{figure}[!ht]
    \centering
    \includegraphics[width=1\linewidth]{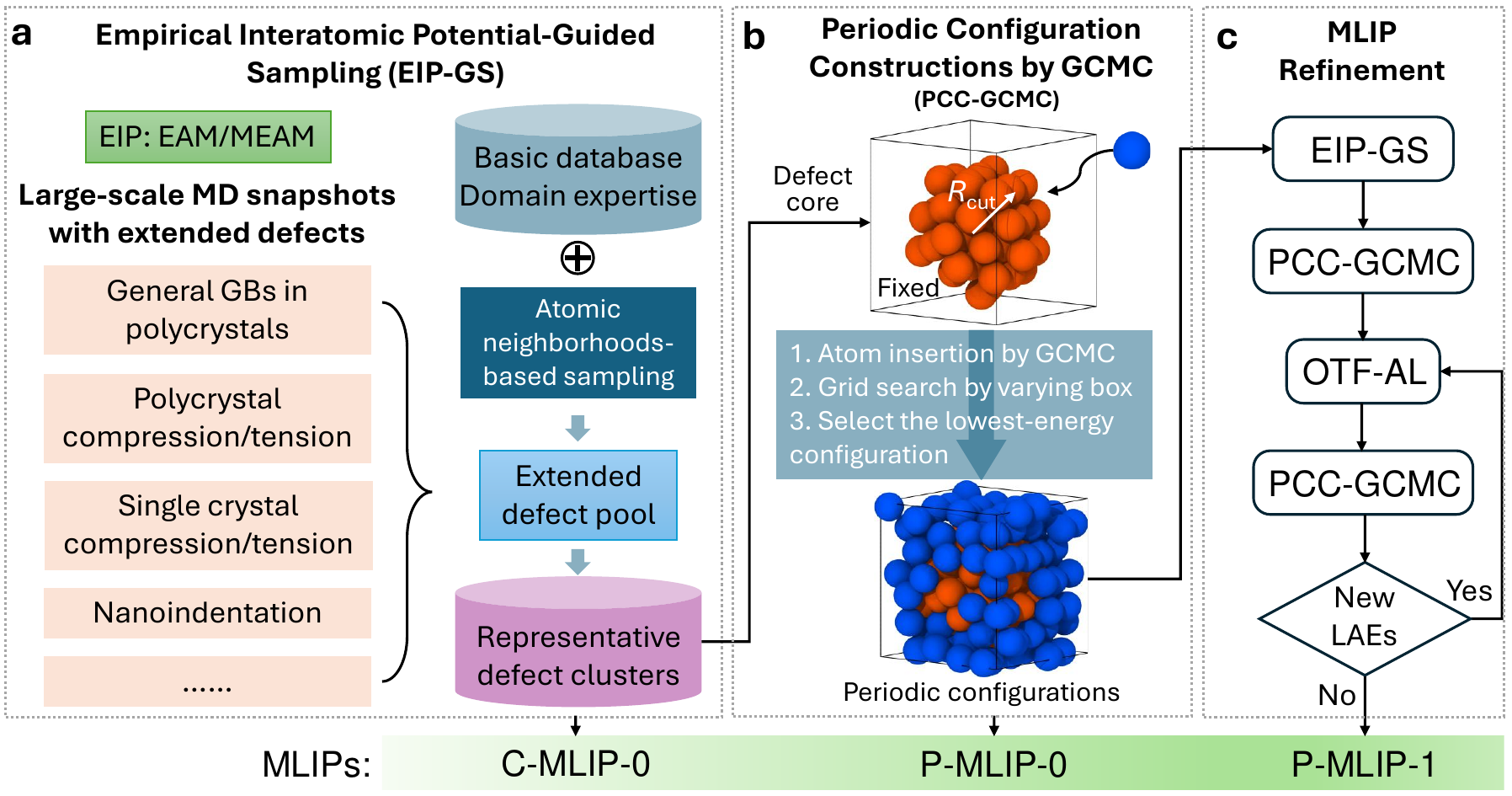}
    \caption{\textbf{The framework of versatile MLIP development for modeling extensive defects.} \textbf{a} Empirical interatomic potential-informed active learning (EIP-GS). \textbf{b} Construction of periodic configurations from non-periodic atomic clusters through precise atom insertion using grand canonical Monte Carlo simulations (PCC-GCMC). \textbf{c} MLIP refinement by combining EIP-GS, on-the-fly (OTF-AL), and PCC-GCMC.}
    \label{fig1}
\end{figure}

\newpage

\begin{figure}[!ht]
    \centering
    \includegraphics[width=1\linewidth]{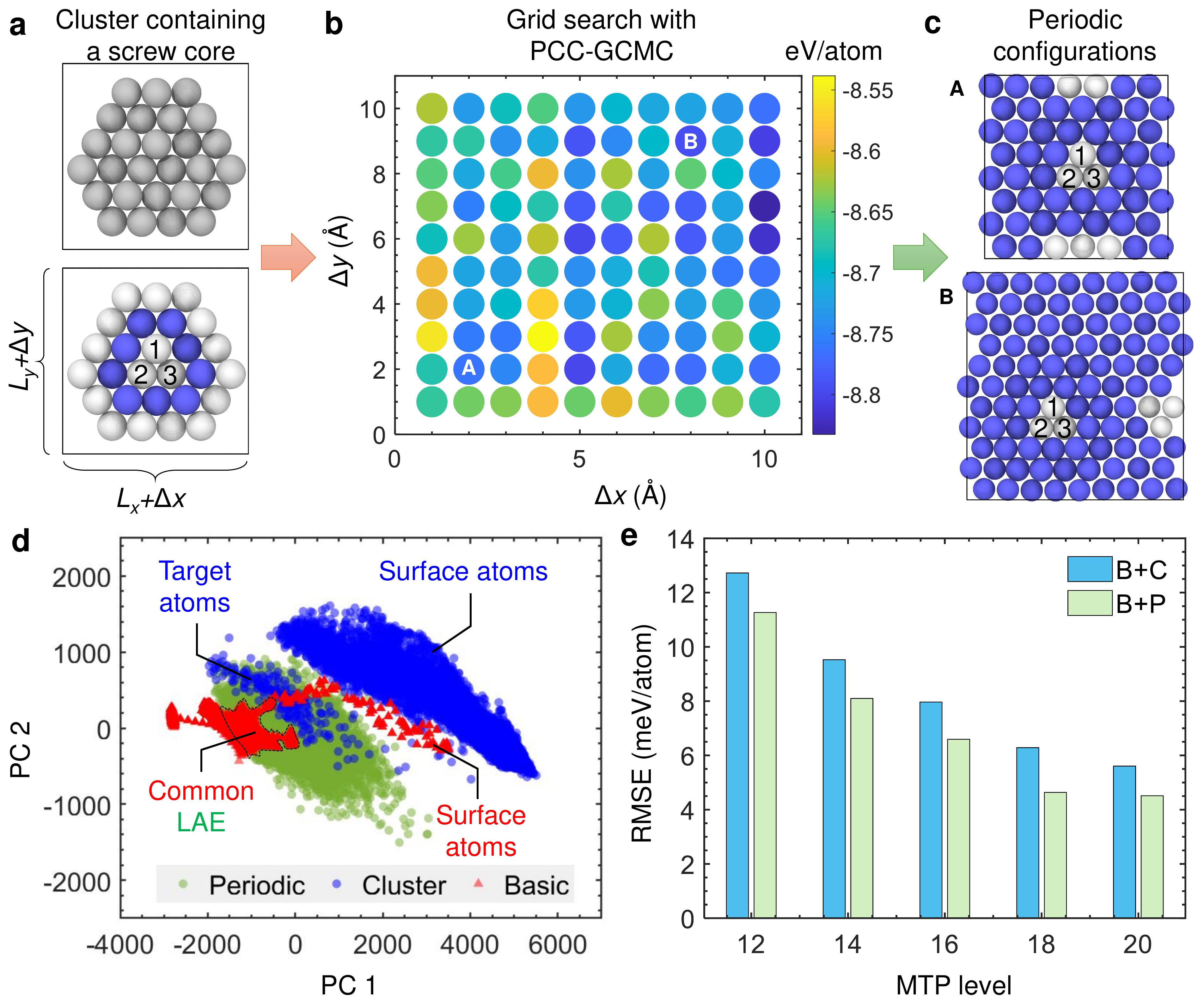}
    \caption{\textbf{Structural analysis of defect pool from EIP-GS.} \textbf{a} Principal component analysis of different datasets. \textbf{b} MTP training performance of two datasets with respect to different levels.}
    \label{fig2}
\end{figure}

\newpage

\begin{figure}[!ht]
    \centering
    \includegraphics[width=0.99\linewidth]{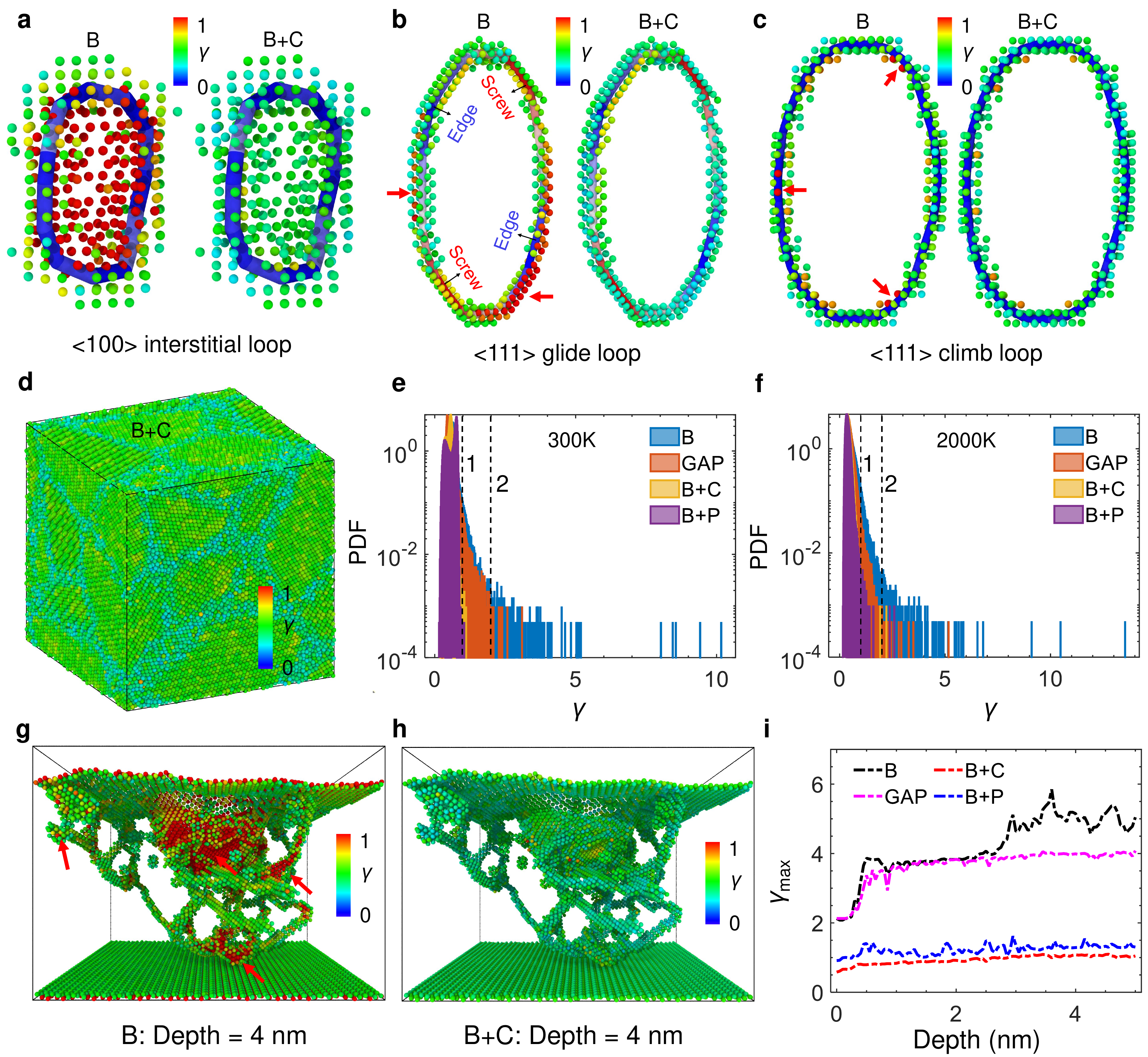}
    \caption{\textbf{Uncertainty quantification of various defects in BCC W based on different dataset.} \textbf{a-c} Three types of dislocation loop, $<100>$ interstitial loop, $<111>$ glide loop, and $<111>$ climb loop. Left panels are results from B, and right panels are results from B+C. Blue lines are edge dislocation segments, and red lines are screw segments. \textbf{d} Polycrystal model. \textbf{e, f} The probability density function (PDF) of extrapolation grade ($\gamma$) at 300 K and 2000 K. \textbf{g, h} Defect structures at the maximum indentation depth from B and B+C. \textbf{i} Evolution of the maximum extrapolation grade ($\gamma_\text{max}$) during the whole indentation process. All atoms in \textbf{a, b, c, d, g, h} are color-coded by $\gamma$. Red arrows in \textbf{b, c, g} indicate the atoms with high uncertainty. A value of $\gamma$ between 0 and 1 indicates interpolation, while a $\gamma$ greater than 1 suggests extrapolation.}
    \label{fig3}
\end{figure}

\newpage

\begin{figure}[!ht]
    \centering
    \includegraphics[width=0.99\linewidth]{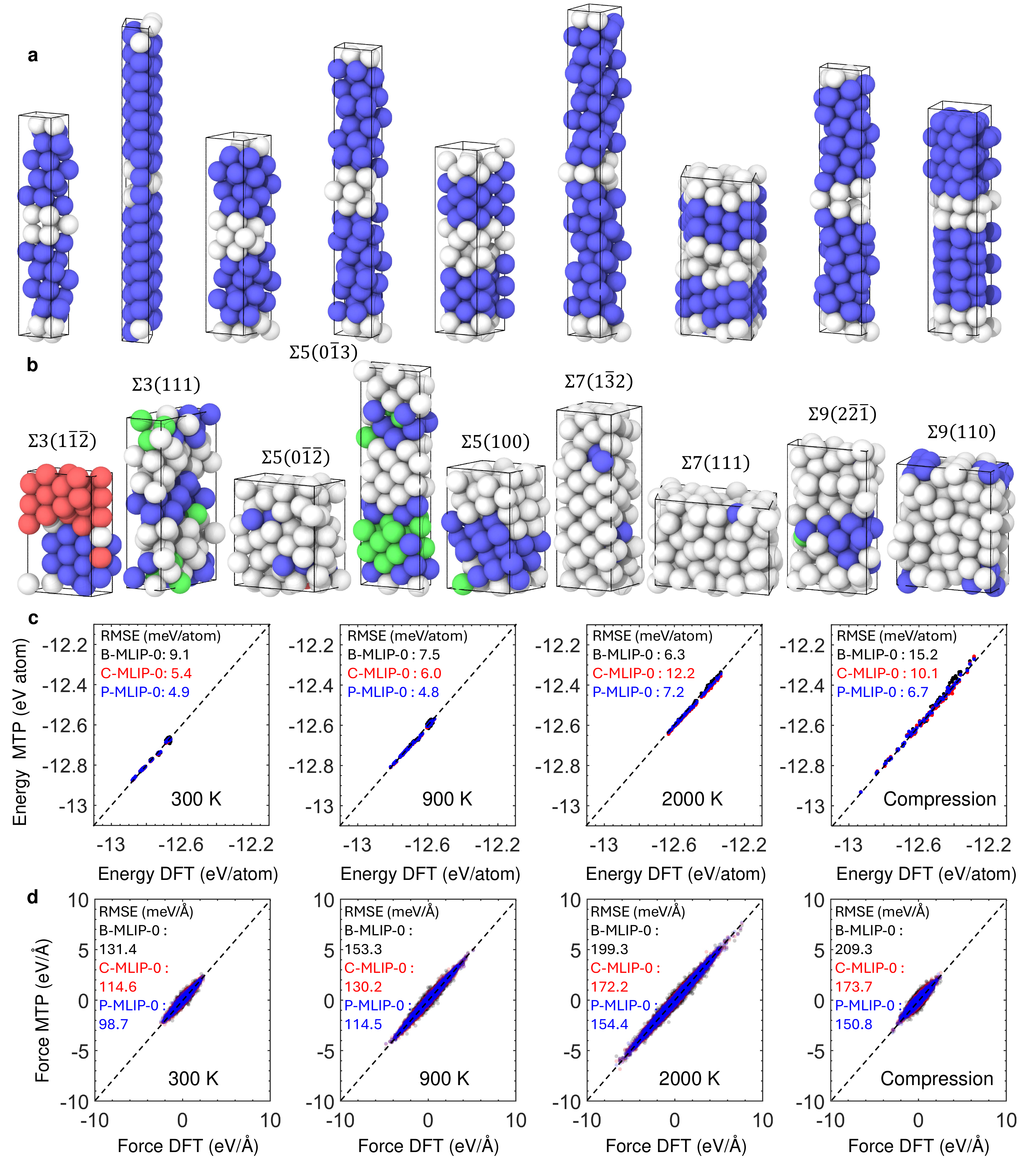}
    \caption{\textbf{Direct validation of MLIPs in nice grain boundaries (GBs) by comparing energy and force from DFT calculations.} \textbf{a} Pristine GB structures. \textbf{b} Compressed GB structures. All atoms in \textbf{a} and \textbf{b} are color-coded by common neighbor analysis. \textbf{c} Energy comparison. \textbf{d} Atomic force comparison.}
    \label{fig4}
\end{figure}

\newpage

\begin{figure}[!ht]
    \centering
    \includegraphics[width=1\linewidth]{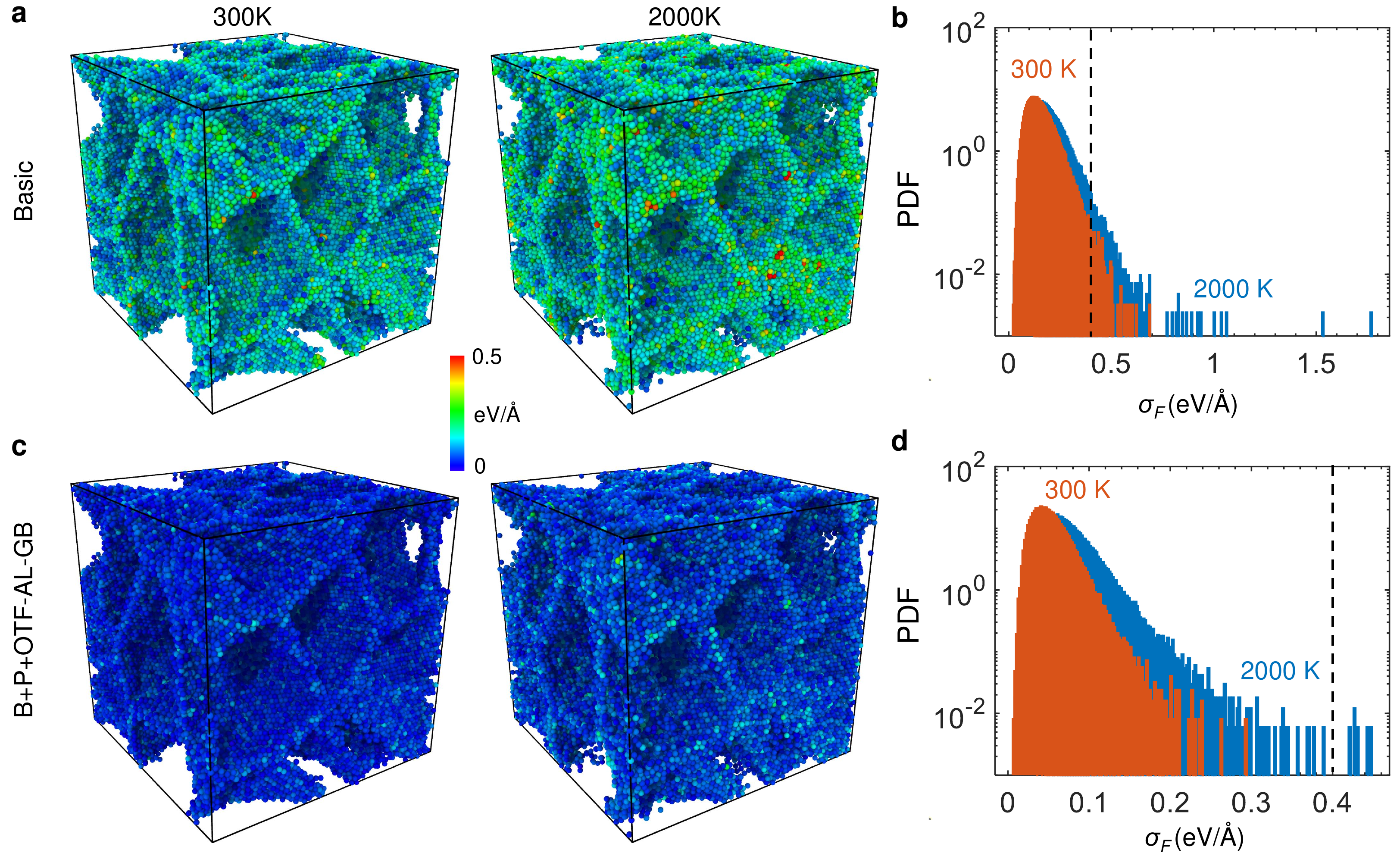}
    \caption{\textbf{Uncertainty quantification based on ensemble learning for a polycrystal.} \textbf{a} Results from the basic dataset. \textbf{b} The probability density function of $\sigma_F$ at temperatures of 300 K and 2000 K, derived from the basic dataset. \textbf{c} Results from the combined dataset (B+P+OTF-GB). \textbf{d} The probability density function of $\sigma_F$ at 300 K and 2000 K based on the combined dataset.}
    \label{fig5}
\end{figure}
\newpage

\begin{figure}[!ht]
    \centering
    \includegraphics[width=1\linewidth]{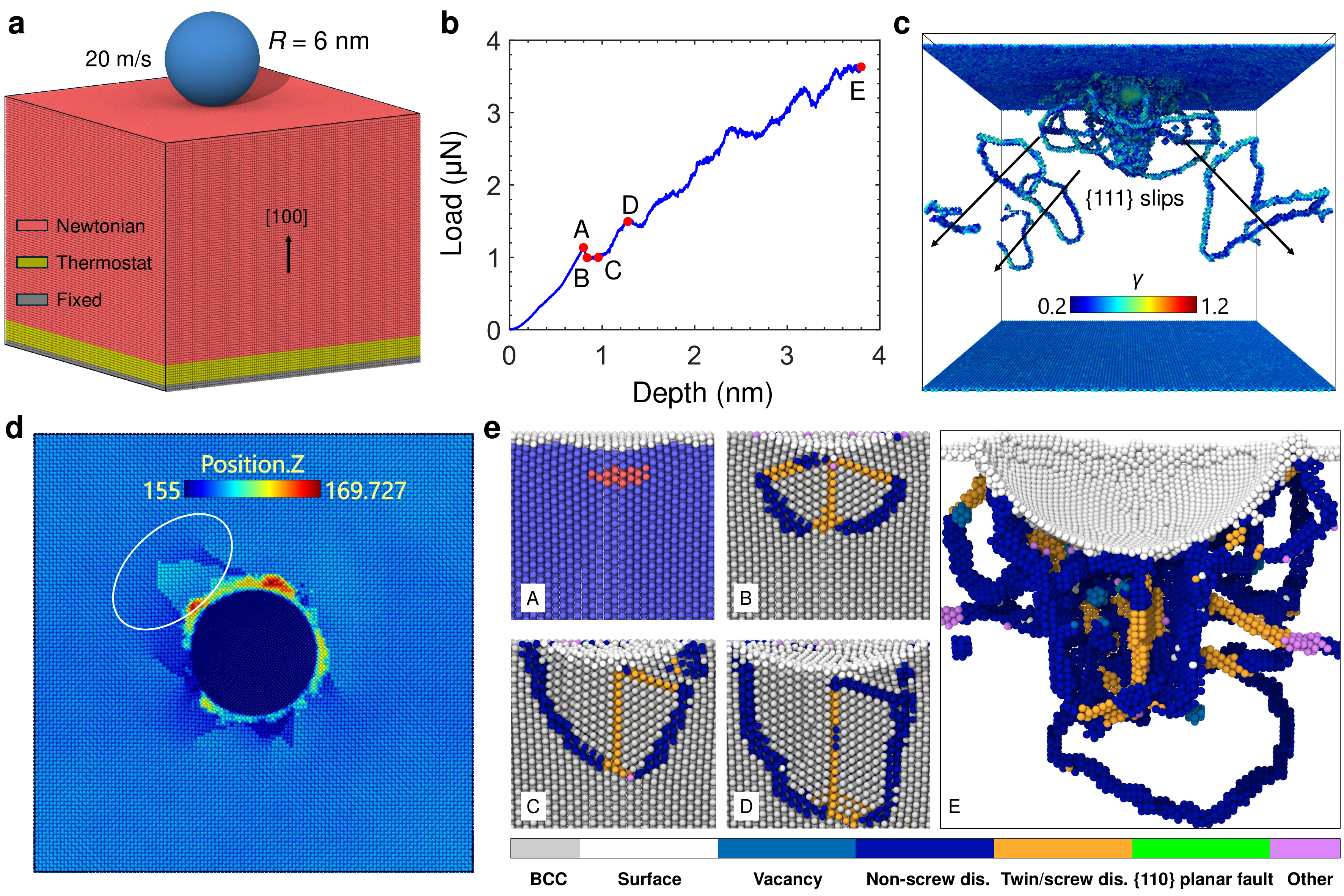}
    \caption{\textbf{Simulation results of nanoindentation in BCC W.} \textbf{a} Simulation setup. \textbf{b} Load-depth curves with five critical points A-E labeled. \textbf{c} The defect pattern at maximum depth, with all atoms color-coded by the extrapolation grade $\gamma$. Arrows indicate the \{111\} slip directions. \textbf{d} Positions of atoms on the top surface; the slip trace is indicated by a circle. \textbf{e} BCC defect analysis (DBA) at five critical points A-E.}
    \label{fig6}
\end{figure}

\newpage

\begin{figure}[!ht]
    \centering
    \includegraphics[width=1\linewidth]{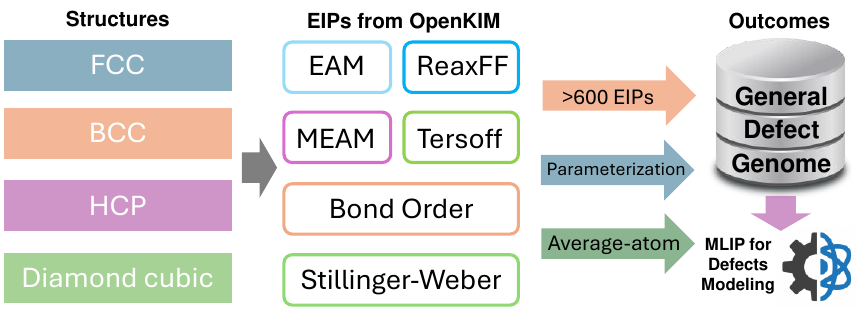}
    \caption{\textbf{Implications of this study on MLIP development for modeling extensive defects in general crystalline materials.}}
   \label{fig7}
\end{figure}

\newpage


\renewcommand{\thefigure}{S\arabic{figure}}
\renewcommand{\thetable}{S\arabic{table}}
\renewcommand{\theequation}{S\arabic{equation}}
\renewcommand{\thepage}{S\arabic{page}}
\setcounter{figure}{0}
\setcounter{table}{0}
\setcounter{equation}{0}
\setcounter{page}{1} 


\begin{center}
\section*{Supplementary Materials for\\ \scititle}

    Fei Shuang$^{\ast}$,
    Kai Liu,
    Yucheng Ji,
    Wei Gao,
    Luca Laurenti,
    Poulumi Dey$^{\ast}$\\
\small$^\ast$Corresponding author. P.dey@tudelft.nl; F.Shuang@tudelft.nl.
\end{center}

\subsubsection*{This PDF file includes:}
Supplementary Note 1-4

\noindent Table S1 and S2

\noindent Figure S1 to S2

\newpage

\subsection*{Supplementary Note 1. LAEs comparison among datasets B+C, B+P and GAP}

The basic dataset contains 678 configurations, which are obtained by domain knowledge. This collection encompasses ground-state structures, configurations deformed under various elastic strains, ab initio molecular dynamics (AIMD) simulations across multiple temperatures, edge and screw dislocations, and simple grain boundaries \cite{GB-2020}. These configurations have been curated from our recent work \cite{Shuang2024}. The GAP dataset is obtained from Ref. \cite{W-GAP-2014}, totaling 9693 configurations. The dataset includes primitive unit cell with varying lattice vectors, vacancy, dislocation quadrupole, unstable stacking fault, surface.

\begin{figure}[!ht]
    \centering
    \includegraphics[width=1\linewidth]{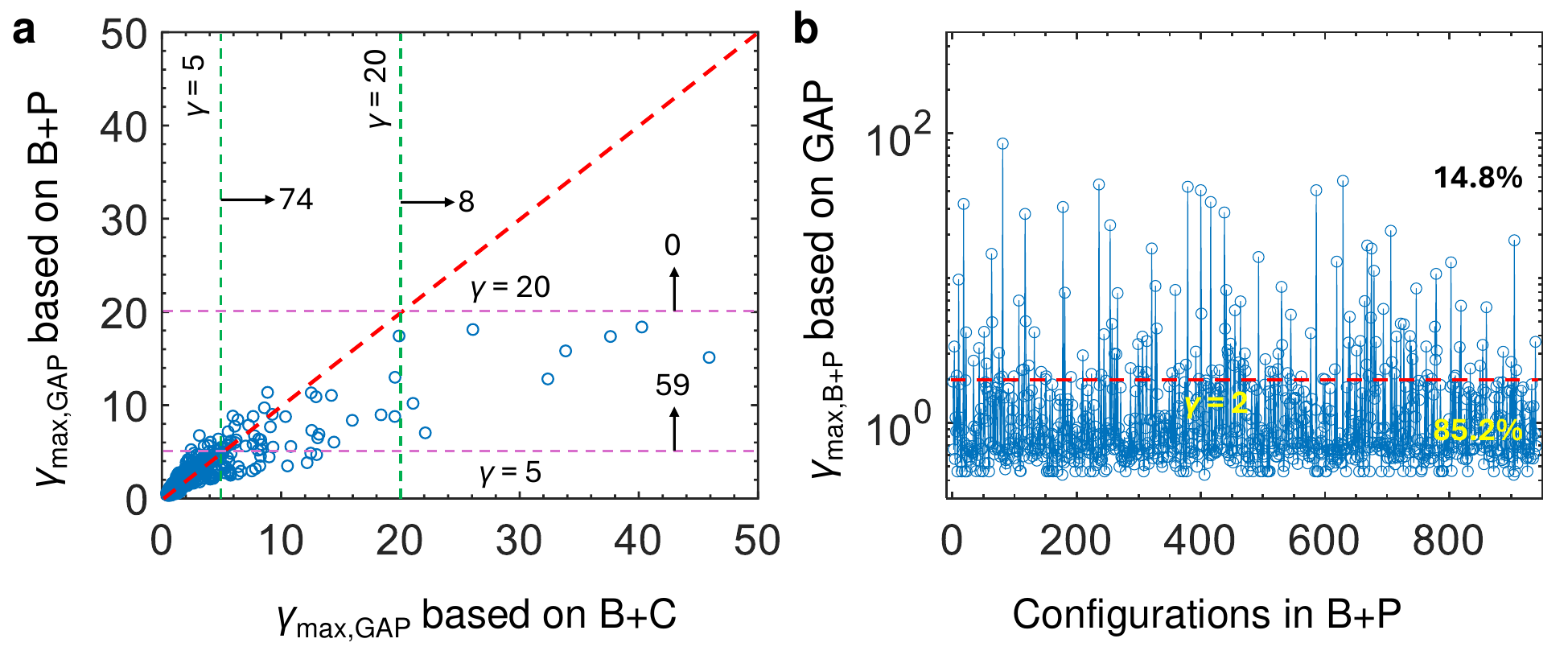}
    \caption{\textbf{Uncertainty quantification based on the extrapolation grade $\gamma$ for the datasets B+C and B+P.} \textbf{a} Comparison of $\gamma_\text{max}$ for the GAP dataset obtained based on B+P and B+C. \textbf{b} $\gamma_\text{max}$ for B+P based on the GAP dataset. A value of $\gamma$ between 0 and 1 indicates interpolation, while a $\gamma$ greater than 1 suggests extrapolation.}
    \label{S1}
\end{figure}

To further highlight the greater diversity of LAE in the B+P dataset compared to B+C, we calculate the extrapolation grade $\gamma$ for configurations within the GAP dataset, evaluated by both B+C and B+P. The comparative results are illustrated in Fig. \ref{S1}a, where we plot the $\gamma$ values for B+C and B+P against each other. Our analysis reveals that the B+P dataset includes 59 configurations with $\gamma$ values exceeding 5, notably, none exceeding 20. In contrast, the B+C dataset shows 74 configurations with $\gamma$ values over 5, including 8 configurations where $\gamma$ surpasses 20. This stark contrast underscores that B+P offers a richer diversity of LAEs than B+C. Interestingly, Fig. \ref{S1}a also indicates that certain LAEs present in the GAP dataset are missing in B+P. These absent LAEs are primarily configurations of dumbbell interstitials, which are particularly relevant to simulations of radiation damage. Nonetheless, incorporating these specific LAEs into our model through EIP-GS and PCC-GCMC would be straightforward. Furthermore, we calculate the extrapolation grade $\gamma$ for configurations within the dataset B+P based on the dataset GAP, as depicted in Fig. \ref{S1}b. It is observed that 14.8\% of configurations within B+P are absent in GAP. This significant absence explains the high uncertainty observed for the dataset GAP in Fig. \ref{fig3}, indicating that the existing GAP dataset and the resultant MLIP are not adequately equipped for general defect modeling of BCC W. This finding underscores the necessity of our study and the development of a more comprehensive dataset such as B+P, which better addresses the complexities of large-scale simulations.

\newpage

\subsection*{Supplementary Note 2. Uncertainty quantification}
We adopt two methodologies for uncertainty quantification. The first utilizes atomic descriptors, specifically the extrapolation grade $\gamma$, as implemented in MLIP-3 \cite{Podryabinkin2023} and similarly applied in ACE \cite{Lysogorskiy2023}, based on the D-optimality criterion and the MaxVol algorithm. A value of $\gamma$ between 0 and 1 indicates interpolation, while a $\gamma$ greater than 1 suggests extrapolation. The second employs ensemble learning, where we train five independent MTPs on the same dataset to compute the standard deviation of atomic force:
\begin{equation}
\sigma_F = \sqrt{\sum_{i \in \{x,y,z\}} \sigma(F_{\alpha,i}^{(1..N)})^2}.
\label{eq:uq}
\end{equation}
Here, $F_{\alpha,i}^{(1..5)}$ is the $i$-th component of the force acting on atom $\alpha$ as obtained by one of the 5 committee members (MLIPs) for a particular configuration; $\sigma$ denotes the standard deviation. IIt should be noted that while both approaches are valuable for uncertainty quantification, they fulfill different objectives. Each method can identify new local atomic environments (LAEs), but the extrapolation grade, $\gamma$, is specifically designed for this purpose and is particularly useful for on-the-fly active learning. However, merely having a good sampling of LAE does not ensure the effectiveness of an MLIP. It is essential to implement a robust MLIP framework that can effectively train these datasets. Therefore, we integrate both methods in our work to leverage their distinct advantages.

\newpage
\subsection*{Supplementary Note 3. MLIP refinement for crack propagation}

\begin{figure}[tp]
    \centering
    \includegraphics[width=1\linewidth]{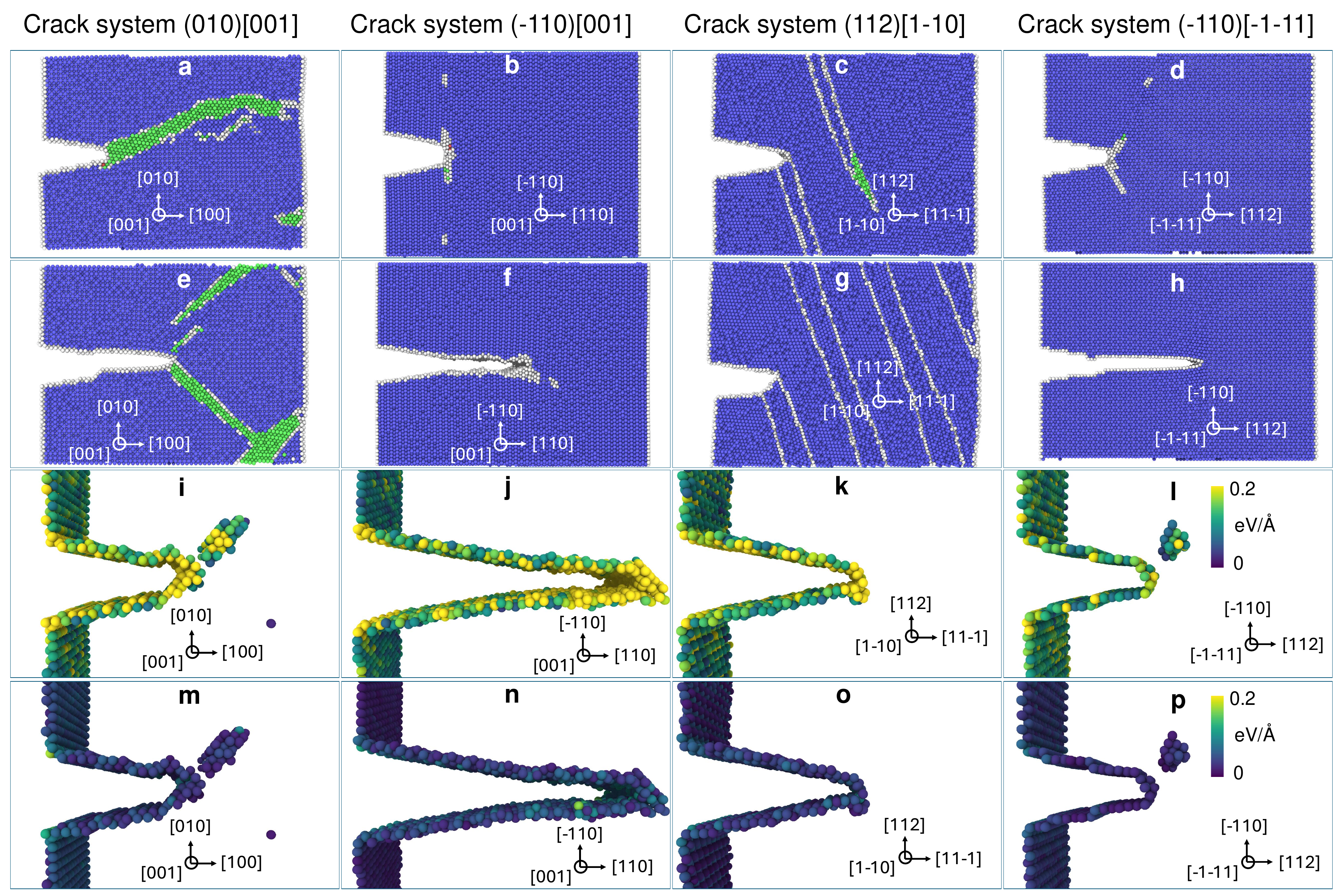}
    \caption{\textbf{EIP-GS and OTF-AL of fracture simulations.} \textbf{a-h} Simulation results of four crack systems using EAM and MEAM potentials. \textbf{i-l} Uncertainty quantification of defects near the crack tip based on MLIPs trained with B+P+OTF-AL-GB. \textbf{m-p} Uncertainty quantification of defects near the crack tip based on MLIPs trained with B+P+OTF-AL-GB+EIP-GS-fracture+OTF-AL-fracture. All atoms in \textbf{a-h} are color-coded by common neighbor analysis: blue atoms have a BCC lattice, green atoms have an FCC lattice, and white atoms have an unrecognized lattice. All atoms in \textbf{i-p} are color-coded by the standard deviation of atomic force.}
    \label{S2}
\end{figure}

In this work, we examine crack propagation in BCC W across four distinct crack systems. Our initial EIP-GS model, referenced in Fig. \ref{fig1}a, did not include fracture simulations. We therefore conducted simulations using both EAM-Zhou \cite{Zhou2004} and MEAM potentials \cite{Hiremath2022}, widely recognized in fracture analysis. We present key snapshots of crack propagation in Figs \ref{S2}a-d using EAM and Figs \ref{S2}e-h using MEAM. MEAM, optimized for superior fracture simulation, exhibits distinct fracture behaviors compared to EAM. For instance, EAM shows a BCC-to-FCC phase transition at the crack tip within the (010)[001] system (Fig. \ref{S2}a), while MEAM primarily displays brittle cleavage (Fig. \ref{S2}e), although a similar phase transition is observed in later stages, potentially influenced by boundary conditions. Additionally, EAM captures dislocation emission in the systems $(\bar{1}10)[001]$ and $(\bar{1}10)[\bar{1}\bar{1}1]$ (Figs \ref{S2}b, d), unlike the brittle cleavage observed with MEAM in these systems (Figs \ref{S2}f, h). Both potentials, however, show deformation twinning at the crack tip in the system $(112)[1\bar{1}0]$ (Figs \ref{S2}c, g). To refine our MLIPs, EIP-GS is applied to simulation outputs from both EAM and MEAM, resulting in the selection of 14 new defect clusters essential for fracture simulations. These clusters are processed using PCC-GCMC to create periodic configurations for further DFT calculations, forming the new dataset, EIP-GS-fracture. Building on this, a new MTP is trained incorporating the datasets B+P+OTF-AL-GB+EIP-GS-fracture. Subsequent OTF-AL for fracture simulations across the four systems leads to the selection of 25 additional configurations. After converting these into periodic configurations via PCC-GCMC and processing them with DFT, they are integrated into the comprehensive datasets, culminating in B+P+OTF-AL-GB+EIP-GS-fracture+OTF-AL-fracture.

To demonstrate the reduction in uncertainty achieved by integrating the EIP-GS-fracture and OTF-AL-fracture datasets, we employ ensemble learning to calculate the standard deviations of atomic forces for critical snapshots during OTF-AL of fracture simulations across four crack systems. These calculations are performed using five independent MTPs trained on datasets B+P+OTF-AL-GB and B+P+OTF-AL-GB+EIP-GS-fracture+OTF-AL-fracture, respectively. The results, presented in Figs \ref{S2}i-p, reveal significant uncertainty at crack tip surfaces and dislocations nucleated from these tips before including crack-related LAEs. However, after incorporating the EIP-GS-fracture and OTF-AL-fracture datasets, the uncertainty for all atoms is substantially reduced, as evidenced by lower standard deviations in atomic forces. This improvement highlights the effectiveness of the combined EIP-GS and PCC-GCMC approaches in adapting to new applications, such as fracture simulations. These findings demonstrate that the final datasets, enriched with a comprehensive defect genome, provide robust capabilities for simulating general plastic deformations and crack propagation in BCC W, offering a solid foundation for accurate modeling across diverse loading configurations and boundary conditions.

\newpage

\subsection*{Supplementary Note 4. Final MLIP training and selection}
We possess a comprehensive final database that aggregates representative LAEs in periodic configurations: B+P+OTF-AL-GB+EIP-GS-fracture+OTF-AL-fracture. Detailed breakdowns of each dataset are presented in Table \ref{TableS1}. Altogether, the database encompasses 1026 configurations and 94,516 atoms. Of the 348 configurations newly incorporated, 276 are obtained through EIP-GS, while only 72 results from on-the-fly AL. This distribution highlights the efficiency of EIP-GS in significantly reducing the reliance on the more resource-intensive OTF-AL, thereby expediting the development of MLIPs.

The MLIPs trained from this database are equipped to simulate nearly all plastic deformations and fracture of BCC W. However, configurations necessary for radiation simulations due to cascade displacement are currently not included. These configurations could easily be integrated from existing databases or generated through EIP-GS and  OTF-AL in future enhancements. We utilize different levels of MTP and a complex ACE potential for training various MLIPs. We particularly favor ACE due to its computational efficiency and the ease of calculating extrapolation grades during simulations in LAMMPS, which greatly facilitates the ongoing assessment. The analysis of basic properties of BCC W, computational efficiency, and training/test performance are detailed in Table \ref{TableS2}. It is observed that the accuracy of MTP models increases with their complexity level, but at the cost of reduced computational efficiency. Specifically, the computational expense of MTP20 is seven times that of MTP12. In contrast, ACE demonstrates accuracy comparable to MTP20 while maintaining efficiency similar to MTP12. Based on these findings, ACE is selected as the optimal model for modeling general defects in BCC W due to its superior balance of accuracy and efficiency.

\newpage

\begin{table}[ht]
\centering
\caption{Datasets generated in this work at different stages.}
\begin{tabular}{lcccccccc}
\hline
& $N_{\text{cfg}}$ & Average atom number & Total atom number \\ \hline
Basic & 678 & 72 & 48816 \\
Initial EIP-GS & 262 & 133 & 34846\\
OTF-AL-GB & 47 & 135 & 6345\\
EIP-GS-fracture & 14 & 118 & 1652\\
OTF-AL-fracture & 25 & 115 & 2875\\
\hline
Total & 1026 & 92 & 94392\\
\hline
\end{tabular}
\label{TableS1}
\end{table}

\newpage
\begin{table}[ht]
\centering
\caption{Comparison of MLIPs and DFT, and training/test errors of MLIPs. The CPU cost is quantified with respect to EAM for each MLIP.}
\small 
\begin{tabular}{lcccccccc}
\hline
& $a_0$ & $C_{11}$ & $C_{12}$ & $C_{44}$ & $\gamma_{110}$ & CPU cost & $E_\text{rmse}$ & $F_\text{rmse}$ \\
& (\AA) & (GPa) & (GPa) & (GPa) & (mJ/m$^2$) & & (meV/atom) & (meV/\AA) \\
\hline
DFT & 3.184 & 520.35 & 199.88 & 142.42 & 1772.74 &  &  &  \\
EAM & 3.165 & 519.93 & 202.17 & 158.96 & 1725.98 & 1 &  &  \\
MTP12 & 3.188 & 461.87 & 196.31 & 116.73 & 1670.87 & 28 & 11.63/11.51 & 212.01/199.51 \\
MTP14 & 3.194 & 480.47 & 204.37 & 109.69 & 1673.98 & 48 & 8.66/6.95 & 196.93/185.71 \\
MTP16 & 3.189 & 462.43 & 185.49 & 127.46 & 1630.25 & 78 & 7.61/6.31 & 182.3/171.84 \\
MTP18 & 3.187 & 537.04 & 214.61 & 140.45 & 1644.52 & 125 & 5.60/4.69 & 155.71/149.05 \\
MTP20 & 3.190 & 530.23 & 195.56 & 146.15 & 1628.74 & 195 & 5.02/4.13 & 141.86/136.46 \\
ACE  & 3.190 & 508.05 & 201.09 & 139.57 & 1641.21 & 30 & 4.08/4.28 & 126.26/160.73 \\
\hline
\end{tabular}
\label{TableS2}
\end{table}

\newpage


\newpage

\newpage

\end{document}